\documentclass[prb,showpacs,twocolumn]{revtex4-1}
\usepackage[dvips]{graphicx}

\begin{document}

\bibliographystyle{apsrev}

\title{Power dependence of pure spin current injection by quantum interference}
\author{Brian A. Ruzicka and Hui Zhao}\email{huizhao@ku.edu}
\affiliation{Department of Physics and Astronomy, The University of Kansas, Lawrence, Kansas 66045, USA}

\begin{abstract}
We investigate the power dependence of pure spin current injection in GaAs bulk and quantum-well samples by a quantum interference and control technique. Spin separation is measured as a function of the relative strength of the two transition pathways driven by two laser pulses. By keeping the relaxation time of the current unchanged, we are able to relate the spin separation to the injected average velocity. We find that the average velocity is determined by the relative strength of the two transitions in the same way as in classical interference. Based on this, we conclude that the density of injected pure spin current increases monotonically with the excitation laser intensities. The experimental results are consistent with theoretical calculations based on Fermi's golden rule.
\end{abstract}


\maketitle

\section{Introduction}

Recently, there has been growing interest in exploring the spin-degree of freedom in semiconductors and its potential use in electronic devices.\cite{s2941488,rmp76323,apl56665} Generation of spin currents is one of the key requirements in spin-based technologies, and has been demonstrated in a number of semiconductor structures. Examples include Si,\cite{n447295} GaAs,\cite{l83203,apl731580,apl812788,n397139,n411770,l94236601,s3092191} InAs,\cite{l88066806,apl90022101} and carbon nanotubes.\cite{n401572,nphys199} Typically, spin currents are produced by dragging spin-polarized carriers with an externally applied electric field. The spin-polarized carriers are obtained either electrically from magnetic contacts or optically by excitation of circularly polarized light. As an intrinsic property of these methods, the spin currents produced are accompanied by charge currents, and are usually called spin-polarized charge currents. Although this type of current can be readily generated and detected electrically, and has been extensively studied, it is desirable to produce spin currents that are not accompanied by charge currents. This type of current is usually called {\it pure} spin current. In the past, pure spin currents, with or without net carrier spin polarizations, have been generated in spin-pump\cite{l91258301} and spin-valve\cite{nphys3197,n448571} configurations, through spin Hall effect,\cite{s3061910,l94047204} circular and spin Galvanic effects,\cite{n417153,SPEChapter,WinklerChapter} and spin dependent phonon scattering.\cite{nphys2609,b75155317} Pure spin currents also exist in pure spin diffusion configurations.\cite{n397139,l94236601,njp9347}

Alternatively, it has been proposed\cite{l855432,b68165348} and demonstrated\cite{l90136603,l90216601,b75075305} that pure spin currents can be injected optically in semiconductors by using a quantum interference and control technique. In this scheme, quantum interference between one-photon and two-photon absorption amplitudes driven by two laser pulses is utilized to control the $k$-space distribution and spin orientation of the excited carriers. By choosing certain polarizations and phases of the laser pulses, one can inject a pure spin current by exciting a certain number of electrons with one spin orientation and an average velocity along one direction, with an equal number of electrons with opposite spin orientation and opposite velocity.\cite{l855432,b68165348,l90136603,l90216601,b75075305} The pure spin current generated by this technique is accompanied by neither a charge current nor a net spin polarization. Furthermore, this noninvasive all-optical technique allows considerable control over the magnitude, sign, duration, and location of the injected current.

One important issue in this technique is the power dependence of current injection. For efficient current injection and precise current modulation in spintronic devices, it is necessary to know how the injected spin current density varies with the intensities of the excitation laser pulses. Fundamentally, study of the power dependence will provide more insights on the mechanisms of the quantum interference process. The density of the injected pure spin current is determined by the electron density and the average velocity of each spin system. The electron density can be readily related to the excitation laser intensities. However, the average velocity, as a result of quantum interference, depends on the {\it relative} strength of the two transition pathways driven by the two laser pulses. Therefore, the overall power dependence of injected current density can be complicated. Indeed, theoretical calculations using different approaches have yielded qualitatively different results.\cite{l855432,b68165348,l95086606,b74165328}

To date, there has been no report on an experimental study of this issue. This is largely due to the fact that, there is no demonstrated techniques available for {\it direct} detection of pure spin currents. Pure spin currents can only be detected indirectly by measuring the spin accumulation caused by the currents.\cite{l90136603,l90216601,b75075305,l96246601,b72201302} Although the spin accumulation can be readily related to a spin separation, these quantities are determined by not only the initial injected current density, but also the relaxation process of the current. The latter is influenced by carrier-carrier scattering, and therefore depends on the carrier density. Hence, the power dependence of spin current injection cannot be obtained by simply measuring the power dependence of the spin separation.

In this paper, we report an experimental study on the power dependence of pure spin current injection by quantum interference. In our approach, we measure the spin accumulation caused by a pure spin current as we vary the relative strength of the two transitions, but keep the current relaxation time constant. This allows us to obtain the power dependence of the injected average velocity. Together with the known power dependence of carrier density, we deduce the power dependence of pure spin current density. Our results are consistent with a perturbation theory based on Fermi's golden rule.\cite{l855432,b68165348}

\section{Average velocity and power dependence}

In the quantum interference and control techniques,\cite{l855432,b68165348} a semiconductor sample is simultaneously illuminated by two phase-locked laser pulses with angular frequencies $\omega$ and 2$\omega$ (Fig.~1a). When $\hbar \omega < E_{\mathrm{g}} <2\hbar \omega$, where $E_{\mathrm{g}}$ is the bandgap of the sample, electrons can be excited from the valence band to the conduction band by one-photon absorption of the $2\omega$ pulse and two-photon absorption of the $\omega$ pulse (vertical arrows in Fig.~1b). Since there are two transition pathways connecting the same initial and final states, quantum interference occurs between the two transition amplitudes. If we control the phases of the transition amplitudes through the phases of the two laser pulses, we can arrange the transition amplitudes to interfere constructively at some $k$-states, but destructively at other $k$-states. This allows us to control the $k$-space distribution of electrons, and inject currents.

\begin{figure}
\includegraphics[width=8.5cm]{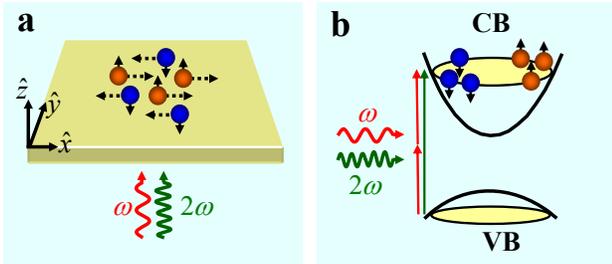}
\caption{(Color online) Quantum interference and control technique to inject pure spin current. Two laser pulses with angular frequencies $\omega$ and $2\omega$ incident to a semiconductor along $+\hat{z}$ direction (panel a). Electrons are excited from the valence band (VB) to the conduction band (CB) via one-photon absorption of the $2\omega$ and two-photon absorption of the $\omega$ pulses (vertical arrows in panel b). When the $\omega$ and $2\omega$ pulses are linearly polarized along $\hat{x}$ and $\hat{y}$ directions, quantum interference causes spin-up electrons (balls with an up-arrow) to be preferentially excited to $k$ states along $+\hat{x}$ direction, with an equal number of spin-down electrons (balls with a down-arrow) to opposite $k$ states (panel b). In real space, electrons with opposite spin orientations move with opposite velocities (horizontal arrows in panel a), forming a pure spin current.}
\end{figure}

Specifically, when both pulses propagate along $+\hat{z}$ direction with the $\omega$ pulse linearly polarized along an arbitrarily chosen $\hat{x}$ direction and the $2\omega$ pulse linearly polarized along the perpendicular $\hat{y}$ direction, spin-up electrons (spin oriented along $+\hat{z}$ direction) with a density of $n_{\uparrow}$ are injected in the conduction band with an average velocity along $+\hat{x}$ direction with a magnitude $v \mathrm{cos}(\Delta \phi)$, where $\Delta \phi=2 \phi _{\omega} - \phi _{2\omega}$ is the relative phase of the two pulses. At the same time, spin-down electrons (spin oriented along $-\hat{z}$ direction) with an equal density $n_{\downarrow}=n_{\uparrow}$ are injected with an opposite average velocity $-v \mathrm{cos}(\Delta \phi)$. Since there are equal number of electrons moving with opposite average velocities (Fig.~1a), there is no net electron transport and no net charge current. However, the spin currents carried by the two spin systems add together, resulting in a pure spin current along $+\hat{x}$ direction. In our experiments, we always choose $\Delta \phi=0$ so that the average velocity is the maximum. The density of the injected pure spin current is\cite{b75075305}
\begin{equation}
\label{PSC_Density}
K=\hbar n v,
\end{equation}
where $n=n_{\uparrow}+n_{\downarrow}$ is the total electron density. We note that a pure spin current carried by the holes in the valanced band is also injected. However, this current is ignored since our detection scheme is set to only detect the electron contribution to the pure spin current (see later discussions).

The nonzero average velocity is the result of an asymmetric distribution of electrons in the conduction band caused by the quantum interferences. Therefore it is determined by the efficiency of the interference. This can be easily understood if one considers the interference of two classical waves. For instance, when two optical beams with the same wavelength and with intensities $I_1$ and $I_2$ interfere, the efficiency of the interference can be described by the contrast of the resulting interference pattern, $A=(I_{\mathrm{MAX}}-I_{\mathrm{MIN}})/(I_{\mathrm{MAX}}+I_{\mathrm{MIN}})$, where $I_{\mathrm{MAX}}$ and $I_{\mathrm{MIN}}$ are maximum and minimum intensities of the interference pattern. It is well known that\cite{BookBornWolf}
\begin{equation}
\label{Classical_interference}
A=\frac{2\sqrt{I_1 I_2}}{I_1+I_2}.
\end{equation}
The most effective interference ($I_{\mathrm{MIN}}=0$) occurs when $I_1=I_2$. In quantum interference, the efficiency of the interference is reflected by the resulting average velocity. Similar to classical interference, the quantum interference efficiency is determined by the relative strength of the two transitions, which can be described by the densities of electrons excited by each pulse acting along, $n_{\omega}$ and $n_{2\omega}$. Remarkably, theoretical calculations based on Fermi's golden rule predict that\cite{l855432,b68165348}
\begin{equation}
\label{Average_Velocity}
v=v_{0}\frac{2\sqrt{n_{\omega} n_{2\omega}}}{n_{\omega}+n_{2\omega}},
\end{equation}
exactly equivalent to the classical interference. The maximum average velocity, $v_{0}$, is achieved when the two transition pathways excite the same electron densities, $n_{\omega}=n_{2\omega}$.

As a direct consequence of this prediction, the density of the pure spin current is, by substituting Eq.~\ref{Average_Velocity} into Eq.~\ref{PSC_Density},
\begin{equation}
\label{Power_Dependence}
K=\hbar v_{0} {2\sqrt{n_{\omega} n_{2\omega}}}.
\end{equation}
We have used the fact that $n=n_{\omega}+n_{2\omega}$. For interband absorption, the excited carrier density is related to the excitation intensity by $n_{2\omega} \propto I_{2\omega}$ for one-photon absorption and  $n_{\omega} \propto I_{\omega}^{2}$ for two-photon absorption.\cite{bookBoyd} Therefore, the power dependence of pure spin current density is\cite{l855432,b68165348}
\begin{equation}
\label{K_Power_Dependence}
K \propto I_{\omega} \sqrt{I_{2\omega}}.
\end{equation}
That is, the injected current density increases monotonically with excitation intensities. However, a recent microscopic many-body model based on semiconductor Bloch equations predicts that the injected current density does not increase monotonically with excitation intensities.\cite{l95086606,b74165328} Therefore, it is desirable to have experimental studies on the power dependence of pure spin current injection by the quantum interference and control technique.

\section{Experimental techniques and procedures}

Figure 2 shows the experimental setup we use to study the power dependence of pure spin current injection. The experiments are performed on both bulk and quantum-well samples of GaAs, at room temperature and 80~K, respectively. Both samples are grown on GaAs substrates along [001] direction. For transmission measurements, the samples are glued on glass substrates and the GaAs substrates are removed by selective chemical etching. Similar results are obtained in both samples. We will first describe the measurements with a 400-nm bulk GaAs sample at room temperature. The measurement on the quantum-well sample will be discussed later. For current injection, the $\omega$ pulse with a central wavelength of 1500~nm and a pulse width of 250~fs is obtained from the signal output of an optical parametric oscillator pumped by a Ti:sapphire laser at 80~MHz. The $2\omega$ pulse is obtained by second harmonic generation from the $\omega$ pulse using a beta barium borate (BBO) crystal. The two pulses are sent through a dichroic interferometer, so that their phases, polarizations, and intensities can be independently controlled. The two pulses are then combined and focused to the sample (Fig.~2a). The $2\omega$ pulse is tightly focused by a microscope objective lens to a spot size $w_0 = 1.4~\mu$m (full width at half maximum). Through one-photon absorption, it excites electrons with a Gaussian spatial profile of the same size. The $\omega$ pulse is focused by the same objective lens to a nominal spot size of $\sqrt{2}w_0$. This is achieved by expanding the $\omega$ beam to $\sqrt{2}$ times bigger than the $2\omega$ beam, considering the spot size is proportional to the wavelength and inverse proportional to the beam size. Since carrier density profile excited by the nonlinear two-photon absorption is $\sqrt{2}$ times narrower than the laser spot, carrier profiles excited by the two pules have the same width $w_0$. This ensures that $n_{\omega}/n_{2\omega}$ is uniform across the whole profile.

\begin{figure}
\includegraphics[width=8.5cm]{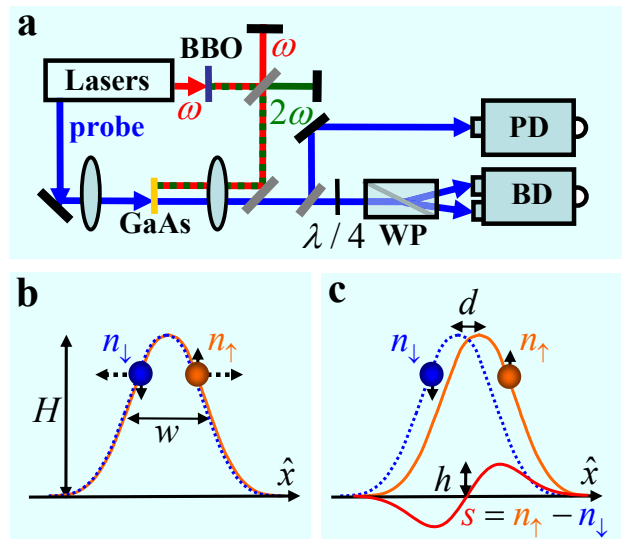}
\caption{(Color online) (a) Experimental setup for injection and detection of pure spin current. See text for details. (b) Upon injection, spin-up and spin-down electrons have identical spatial profiles (Gaussian shaped solid and dotted curves) with a height of $H$ and a width of $w$ (full width at half maximum), but with opposite average velocities (horizontal arrows). (c) After the transport, the two profiles separate by a distance $d$ (spin separation). The resulting spin density ($s$) has a derivative-like profile, with a height $h$ related to $d$ by Eq.~\ref{Gauss_derivative}.}
\end{figure}

Upon injection, the density profiles of the spin-up and spin-down electrons overlap in space, but with average velocities along $+\hat{x}$ and $-\hat{x}$ directions, respectively (Fig.~2b). Therefore, the two profiles move oppositely. Since there is no driving force, the injected average velocity decays rapidly due to the scattering of electrons with phonons and other carriers. Under typical conditions, the relaxation time is less than 1~ps.\cite{b75075305} Figure~2c illustrates the situation right after the current relaxation. The two profiles are separated by a distance $d$. Clearly, the separation of the two profiles results in the accumulation of spin-up and spin-down electrons on opposite sides of the profiles. Due to the short lifetime of the currents, the final spin separation is usually much smaller than the size of the profiles. Therefore, the spin density $s \equiv n_{\uparrow}-n_{\downarrow}$ has a spatial profile similar to the derivative of Gaussian profile, as shown as the solid derivative-like curve in Fig.~2c. Quantitatively, the spin separation is proportional to the height of the spin density profile, $h$, as\cite{b72201302}
\begin{equation}
\label{Gauss_derivative}
d=0.707w\frac{h}{H},
\end{equation}
where $H$ and $w$ are the height and width (full width at half maximum) of the Gaussian density profile of each spin system. Hence, by measuring the profiles of electron and spin densities, we can deduce the spin separation.

The electrons and the spin densities are measured by a pump-probe technique.\cite{b78045314} The linearly polarized, 200-fs probe pulse with a central wavelength of 850~nm is obtained by second harmonic generation of the idler output of the optical parametric oscillator. It is focused to the sample to a spot size of about 1.4~$\mu$m from the other side of the sample (Fig.~2a). A portion of the transmitted probe pulse is reflected to a photodiode (PD) connected to a lock-in amplifier. It is used to measure a differential transmission, $\Delta T/T_0 \equiv [T(n)-T_0]/T_0$, i.e., the normalized difference between transmission in the presence of carriers [$T(n)$] and without them [$T_0$]. Under our experimental conditions, we verify that $\Delta T/T_0 \propto n$. Spin density is measured by the same probe pulse. The linearly polarized probe is composed of two circular components. Due to spin-selection rules, each component preferentially senses electrons with one spin orientation.\cite{opticalorientation} A portion of the transmitted probe pulse is sent through a quarter-wave plate ($\lambda /4$). The two circular components are converted to two orthogonal linear polarizations. A Wollaston prism (WP) is used to spatially separate the two components and send them to two photodiodes of a balanced detector (BD). The output voltage of the balanced detector is proportional to the difference between the differential transmissions of the two circular components, $(\Delta T^+ -\Delta T^-)/T_0$, which is proportional to the spin density $s$.\cite{b72201302,l96246601,b75075305} This output is measured by a lock-in amplifier.

In all of the measurements, the probe pulse is arranged to arrive at the sample 3~ps later than the pump pulses. This probe delay is chosen for the following reasons. First, this time is long enough for the spin transport process to complete, so that the {\it final} spin separation is measured. Second, this probe delay time is longer than the spin relaxation time of holes, which has been reported to be much shorter than 1 ps in bulk GaAs.\cite{l89146601} This ensures that we only detect spin current carried by electrons, since the effect of hole spin current doesn't persist for that long of a time. Finally, this delay time is much shorter than the spin relaxation time and lifetime of electrons that are both longer than 100 ps. Therefore, the spin density caused by the current doesn't decay significantly.\cite{b75075305}

\section{Results and discussions}

Figure~3 shows an example of the measured electron and spin density profiles measured by scanning the probe spot along $\hat{x}$ direction. In this measurement, the energy fluence of the two pump pulses are adjusted to produce electron densities of $n_{\omega}=n_{{2\omega}}=1.25~ \times 10^{17} \mathrm{cm}^{-3}$ at the center of the profile. The Gaussian profile of the electron density (squares) is consistent with the size and shape of the laser spots. We note that the broadening of the profile due to carrier diffusion is negligible on this time scale.\cite{spindiffusionprb} The spin density profile is shown as the circles. The solid line is a fit with the derivative of a Gaussian function. From these profiles, we deduce a spin separation of 44~nm.

\begin{figure}
\includegraphics[width=8.5cm]{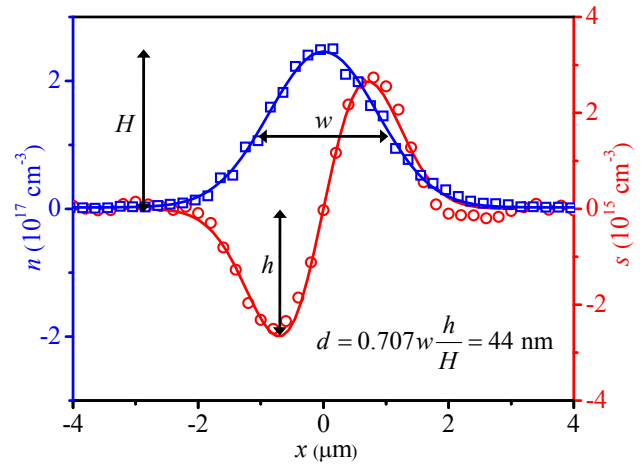}
\caption{(Color online) Spatial profiles of the total electron density $n=n_{\uparrow}+n_{\downarrow}$ (squares, left axis) injected in the GaAs bulk sample at room temperature and the spin density $s=n_{\uparrow}-n_{\downarrow}$ (circles, right axis) that resulted from spin transport. $x=0$ is defined as when the pump and probe spots overlap. A spin separation of 44~nm can be deduced from these profiles.}
\end{figure}

The procedure summarized in Fig.~3 is repeated with various combinations of $n_{\omega}$ and $n_{2\omega}$, by adjusting the energy fluences of the two pump pulses, however keeping the total electron density constant. The spin separations deduced from these measurements are plotted as a function of $n_{\omega}/n_{2\omega}$ in Fig.~4 (circles). The maximum spin separation occurs when $n_{\omega}=n_{2\omega}$. Furthermore, the whole set of measurements is repeated with other total electron densities (1.4 and 5.5~$\times 10^{17} \mathrm{cm}^{-3}$, respectively). Similar results are obtained in both sets of measurements, as shown in Fig.~4 (squares and triangles).

The spin separation is determined by the initial average velocity injected and the sequential spin transport process. After injection, the spin transport is controlled by scattering of electrons with phonons and other carriers. By keeping the total electron density and the lattice temperature constant, the relaxation process of the current is not expected to change as we vary  $n_{\omega}/n_{2\omega}$ in each set of measurements. Therefore, although we are not able to determine the exact value of the injected average velocity, its {\it dependence} on $n_{\omega}/n_{2\omega}$ is the same as that of the spin separation. This allows us to compare our experimental results with theory. We fit each data set with Eq.~\ref{Average_Velocity}, allowing a constant factor as the only adjustable parameter. In Fig.~4, each data set has been scaled by multiplying a factor that is given in the caption of the figure, so that all the fitted curves overlap (solid line). Clearly, our experimental results are consistent with Eq.~\ref{Average_Velocity}, which is based on Fermi's golden rule.\cite{l855432,b68165348} We note that the spin separation doesn't change significantly with the total electron density. This is reasonable since at room temperature and with moderate electron densities, the current relaxation is likely to be controlled by phonon scattering.

Since no assumption is needed in deducing Eq.~\ref{K_Power_Dependence} from Eq.~\ref{Average_Velocity}, the consistency between the symbols and the solid line in Fig.~4 verifies the power dependence of the pure spin current injection by quantum interference and control techniques, Eq.~\ref{K_Power_Dependence}, as predicted by the model based on Fermi's golden rule.\cite{l855432,b68165348} Furthermore, we extend the experiment to a quantum-well sample containing 10 periods of 14-nm GaAs quantum wells sandwiched by 14-nm AlGaAs barriers. The sample is cooled to 80~K. The measurement is performed in a similar fashion, with a total electron density of 1.0~$\times 10^{17} \mathrm{cm}^{-3}$. The only difference worth mentioning is that, the probe beam in this measurement is obtained from the Ti:sapphire laser and tuned to 807~nm ($1s$ heavy hole exciton resonance). The measured spin separations are plotted in Fig.~4 (diamonds), after scaled by multiplying a factor of 0.28, along with results from the bulk sample at room temperature. The spin separations at 80~K are about three times larger than those at room temperature. This can be easily understood since the phonon scattering is suppressed at 80~K. However, the same dependence on $n_{\omega}/n_{2\omega}$ is obtained. This gives us more confidence that the spin separation we measured is indeed strictly proportional to the average velocity and is not influenced by relaxation of the current in each set of measurements.

\begin{figure}
\includegraphics[width=8.5cm]{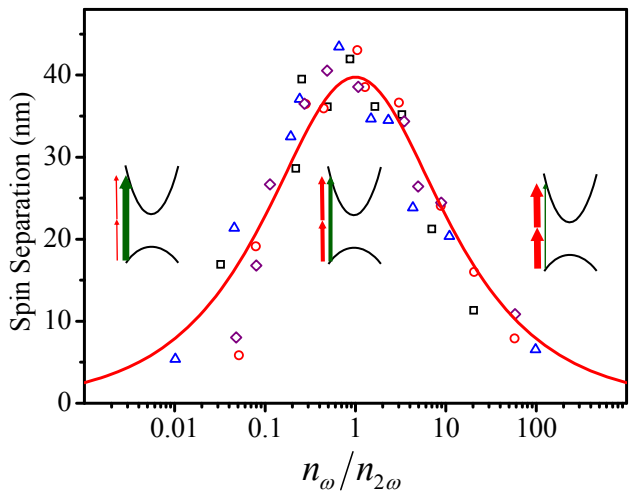}
\caption{(Color online) Spin separation measured as a function of $n_{\omega}/n_{2\omega}$ by using the procedure summarized in Fig.~3. In each set of measurements, the total electrons density is kept constant when $n_{\omega}/n_{2\omega}$ is varied. The squares, circles and triangles show data measured from the bulk sample at room temperature, with the total electron density at the center of the profile of 1.4, 2.5, and 5.5~$\times 10^{17} \mathrm{cm}^{-3}$, respectively. The diamonds represent data measured from the quantum-well sample at 80~K, with the total electron density of 1.0~$\times 10^{17} \mathrm{cm}^{-3}$. The data sets shown as squares, triangles and diamonds are scaled by multiplying factors of 1.30, 1.42, and 0.28, respectively, so that all the fitted curves by Eq.~\ref{Average_Velocity} overlap, shown as the solid line.}
\end{figure}

Previously, a non-monotonic power dependence of current injection by quantum interference was predicted by a microscopic many-body theory.\cite{l95086606,b74165328} It was shown that, when the $I_{2\omega}$ is kept constant, the injected current density increases linearly with $I_{\omega}$, as predicted by the perturbation theory,\cite{l855432,b68165348} but only for low $I_{\omega}$. Significant derivation from linear dependence appears when $I_{\omega}/I_{2\omega}$ approaches nine. When $I_{\omega}/I_{2\omega}$ is about sixteen, the deviation is more than 50\%. Furthermore, when $I_{\omega}/I_{2\omega}$ is further increased, the current density {\it decreases}. In our experiment, the power dependence is studied in the range of $10^{-2}<n_{\omega}/n_{2\omega}<10^{2}$. With a one-photon absorption coefficient of $10^{4} \mathrm{cm^{-1}}$ and a two-photon absorption coefficient of $10 \mathrm{GW cm^{-2}}$,\cite{ol32668} this range corresponds to $10<I_{\omega}/I_{2\omega}<10^{4}$. We note that the previous calculation is based on quantum wires, since two dimensional calculation is too time-consuming.\cite{l95086606,b74165328} Although qualitatively similar results were obtained in quantum wells,\cite{l95086606,b74165328} significant quantitative differences can be expected. Therefore, a direct quantitative comparison of our experimental results on bulk and quantum-well samples with the theoretical results on quantum wires is irrelevant. Our experiment demonstrates that the simple perturbation theory\cite{l855432,b68165348} is adequate in describing spin current injection by quantum interference under typical experimental conditions used in this study and previous studies of this type.\cite{l90136603,l90216601,b75075305} It would be interesting to experimentally explore non-perturbation regimes of current injection by quantum interference, and to theoretically study non-perturbation effect on current injection in bulk and quantum wells under typical conditions.

Finally, we note that our method can {\it not} be readily generalized to study the power dependence of pure {\it charge} current injection by quantum interference.\cite{l761703,l78306} A same power dependence of charge current injection has been predicted by theory.\cite{l855432,b68165348} However, unlike pure spin current, in charge current the transport gives rise to a space charge field that significantly changes the dynamics of the transport.\cite{jap103053510} Since the space charge field cannot be isolated from the average velocity, it is difficult to relate the electron accumulation to the initial average velocity.\cite{jap103053510} Nevertheless, the same power dependence of charge current injection has indeed been observed in GaAs, silicon, germanium, and carbon nanotubes by measuring THz emission from the samples.\cite{nphys3632,b77085201,nl81586} We suggest that the consistency between our experiment and the THz-based charge current experiments\cite{nphys3632,b77085201,nl81586} is an indication that the THz signal detected in those experiments\cite{nphys3632,b77085201,nl81586} is proportional to the charge current density initially injected, and is not influenced by the sequential charge transport dynamics.

\section{Summary}

We have studied the power dependence of pure spin current injection in GaAs bulk and quantum-well samples by the quantum interference and control technique. Although we cannot directly measure the density of injected pure spin current, the spin separation caused by the current can be deduced from the spin density measured using the pump-probe technique. The spin separation is measured as a function of the ratio of the electron densities excited by the two transitions, $n_{\omega}/n_{2\omega}$, while the sum of the two densities is kept constant. We found that the spin separation reaches maximum when the ratio is one. Since the total electron density and lattice temperature are unchanged as the $n_{\omega}/n_{2\omega}$ is varied, the spin separation is proportional to the initially injected average velocity by quantum interference. We found that under our experimental conditions, the average velocity is determined by $n_{\omega}/n_{2\omega}$ in the same way as classical interference, as predicted by the model based on Fermi's golden rule.\cite{l855432,b68165348} As a consequence, the density of the injected pure spin current increases monotonically with the excitation intensities.

\section*{Acknowledgement}

We acknowledge John Prineas of University of Iowa for providing us with high-quality GaAs bulk and quantum-well samples, and financial support of the General Research Fund of The University of Kansas.

\end{document}